\documentclass[aps,pre,groupedaddress,twocolumn,10pt]{revtex4-1}
\usepackage{graphicx}
\usepackage{epstopdf}
\usepackage[english]{babel}
\usepackage[T1]{fontenc}
\usepackage{verbatim}
\usepackage{float}

\usepackage{makecell}
\usepackage{color}
\usepackage{multirow}
\usepackage{slashbox}
\usepackage{booktabs}
\usepackage{amsmath}
\usepackage{amssymb}
\usepackage{amsthm}
\usepackage{bbm}
\usepackage{hyperref}
\usepackage[T1]{fontenc}
\usepackage[usenames,dvipsnames]{xcolor}
\hypersetup{colorlinks=true, linkcolor=blue, urlcolor=blue!50!black, citecolor=blue!50!black}%BrickRed

\usepackage[normalem]{ulem}

\renewcommand{\imath}[0]{\mathsf{i}}

\usepackage[nodayofweek,level]{datetime}

\begin{document}

\title{Critical condition for electrowetting-induced detachment of a droplet from a curved surface}

\author{Ke Xiao and Chen-Xu Wu$^{*}$}

\affiliation{Department of Physics, School of Physical Science and Technology, Xiamen University, Xiamen 361005, People's Republic of China}
\email{cxwu@xmu.edu.cn}

\begin{abstract}

Based on energy conservation, we derive a critical condition theoretically for electrowetting-induced droplet detachment from a hydrophobic curved surface. Phase diagrams are constructed in terms of droplet volume, viscosity, Ohnesorge number, friction coefficient at contact line, surface curvature, surface wettability and electrowetting number. The deduced critical condition offers a general and quantitative prediction on when the detachment occurs, a criterion enabling us to gain more insights into how to accurately manipulate the electrowetting-induced detachment of an aqueous droplet on a curved surface.
The results obtained in this paper also imply that the detachable regimes of the phase diagrams can be enlarged through increasing droplet volume and surface curvature, and reducing liquid viscosity, friction coefficient, Ohnesorge number and wettability of substrates.

\end{abstract}
\date{\today}

\maketitle

%%%MAIN TEXT%%%%
Directly harnessing detachment and interfacial geometric shape of droplet on a flat substrate has sparked an increasing amount
of interest because of its significance in fundamental scientific understanding and engineering and technological applications, including self-cleaning~\cite{K.M.Wisdom2013,L.Yao2014}, anti-icing/dew~\cite{Q.L.Zhang2013,J.B.Boreyko2009}, inkjet/soft printing~\cite{T.Boland2006,L.B.Zhang2015}, fast response displays~\cite{R.A.Hayes2003}, fast optical imaging~\cite{C.L.Hao2014}, optical devices~\cite{S.Kuiper2004,J.Heikenfeld2005}, and novel digital microfluidic devices~\cite{K.Choi2012,S.JunLee2012,J.Hong2015,P.G.Zhu2017}.
Except for spreading and evaporation when a sessile droplet initially rests on solid substrate, it is also able to depart from the substrate with some external stimulation.
Generally, there are various approaches to generating detachment of droplet from the substrate, such as electrowetting (EW)-induced detachment~\cite{S.JunLee2012,S.J.Lee2014,A.Cavalli2016,Z.T.Wang2017,B.Traipattanakul2017,Q.Vo2019,K.X.Zhang2019,C.T.Burkhart2020,Q.G.Wang2020}, coalescence-induced jumping~\cite{Y.Nam2013,R.Enright2014,T.Mouterde2017,H.Vahabi2018}, impact-caused bouncing~\cite{Y.H.Yeong2014,B.Y.Zhao2017,A.Bordbar2018,M.Abolghasemibizaki2019,Olinka2020}, acoustic wave actuated bouncing~\cite{M.H.Biroun2020,M.H.Biroun2020PRApp}, light-triggered bouncing~\cite{W.Li2020}, and Laplace pressure driven jumping~\cite{X.Yan2020}.
Among them, EW has become a prevalent technique due to its advantages of easy fabrication and manipulation, adaption to various geometries, short response time, little power consumption and high reversibility~\cite{F.Mugele2005,W.C.Nelson2012}.
It is widely known that if an electric voltage is applied between the droplet and the substrate, a droplet settling on a flat substrate will alter its apparent wettability (contact angle), a phenomenon referred to electrowetting-on-dielectric (EWOD)~\cite{F.Mugele2005,L.Q.Chen2014}.
Over the past two decades, numerous efforts using experiment~\cite{S.JunLee2012,S.J.Lee2014,A.Cavalli2016,Z.T.Wang2017,B.Traipattanakul2017,Q.Vo2019,C.T.Burkhart2020,Q.G.Wang2020}, theoretical modeling~\cite{P.Birbarah2015,A.Cavalli2016,K.X.Zhang2019}, and numerical simulation~\cite{K.A.Raman2016,A.Merdasi2019june,A.Merdasi2019oct} have been devoted to getting a better understanding of the dynamics of droplet detachment induced by EW.

Experimentally, it has been found that a variety of factors, such as viscosity~\cite{Q.Vo2019,Q.Vo2018PRE,Q.G.Wang2020}, surface wettability~\cite{A.Cavalli2016}, contact line friction~\cite{Q.Vo2019,Q.Vo2018PRE,Q.G.Wang2020}, and droplet volume~\cite{C.T.Burkhart2020}, play a prime role in determining the dynamics of droplet detachment from a flat solid substrate induced by EW.
For example, by taking surface energy difference, viscous dissipation and stored energy into account, a critical condition is proposed to determine the detachable and nondetachable regimes~\cite{Q.Vo2019}. Besides, a critical equation for threshold value of applied electric voltage which triggers droplet detachment is built and further confirmed by experimental data~\cite{Q.G.Wang2020}.
In addition, theoretical modeling and numerical simulation also provide a useful complement to understanding the underlying mechanism of droplet detachment from a flat substrate.
By combining experiments and numerical simulations, the process of EW-induced ejection of a droplet from a flat substrate has been quantitatively analyzed, and the effect of applied voltage and intrinsic contact angle on the energy conversion efficiency have been further discussed~\cite{A.Cavalli2016}. A similar investigation reveals that the droplet detachment velocity can be predicted accurately by studying the dynamic process of ejection quantitatively~\cite{K.X.Zhang2019}.

Despite of the fact that dynamic process of EW-induced droplet detachment has been widely studied either via experimental or theoretical approaches, the efforts are all focused on flat solid substrates. Numerous standing questions still remain open concerning the detachment mechanism of an aqueous droplet on a curved surface induced by the same EW effect. A detailed investigation of the relations between the threshold of the applied electric voltage and the system parameters, e.g., droplet volume, liquid properties, surface curvature, and Young's contact angle etc., on curved surfaces, is needed to broaden the current understanding of droplet detachment.

In this paper, we perform a systematic theoretical study on EW-induced droplet detachment from a curved surface by building expressions for interfacial energy, energy of dissipation, and accumulated energy. A criterion equation for EW-induced droplet detachment, which is able to return to that for a flat surface, is theoretically derived. To further explore the underlying dependence of the threshold of the applied electric voltage on droplet volume, liquid properties, surface curvature, and Young's contact angle, phase diagrams are also constructed.

\begin{figure}[htp]
  \includegraphics[width=\linewidth,keepaspectratio]{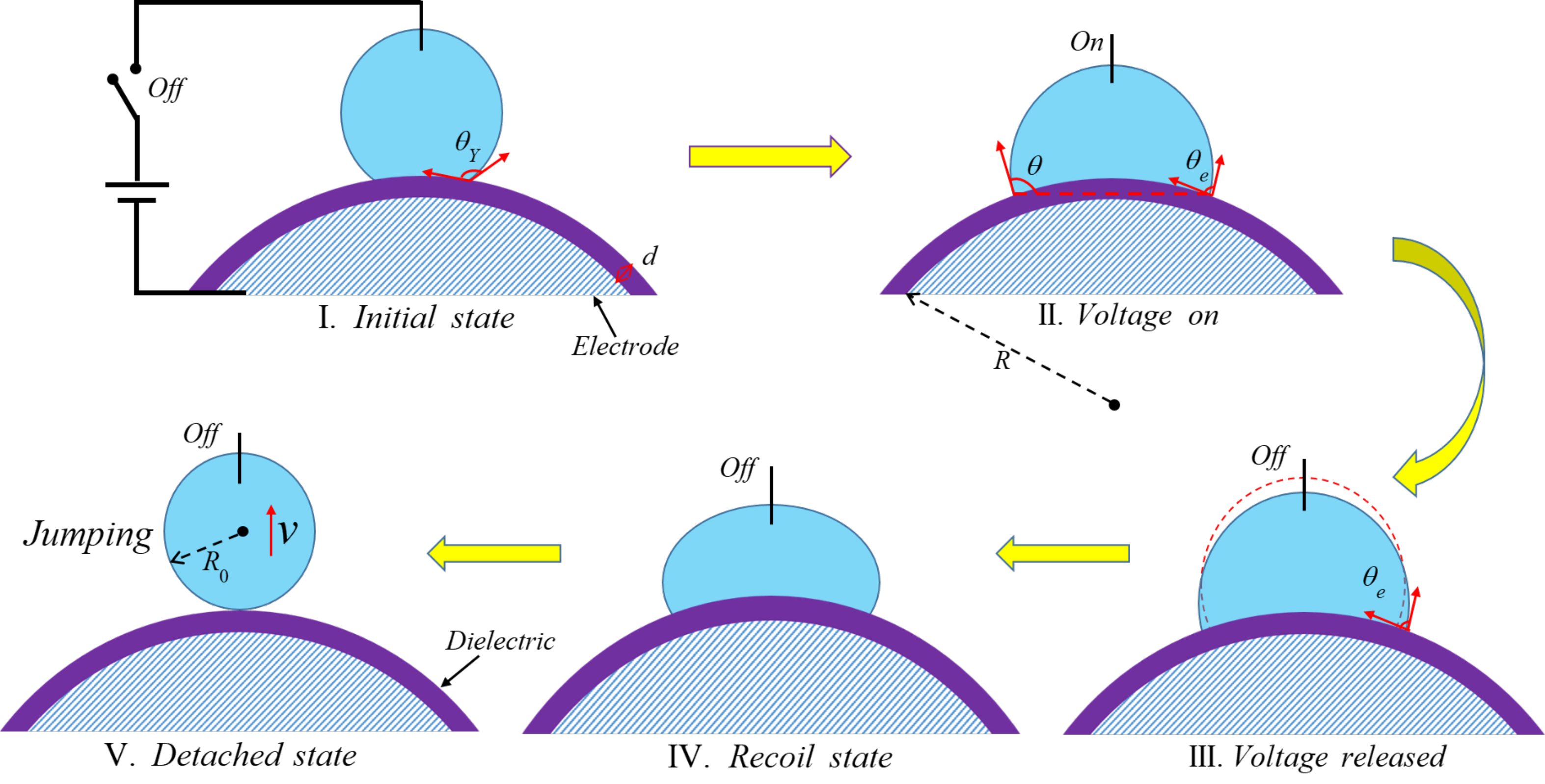}
  \caption{(Color online) Schematic pictures of different scenarios of electrically actuated droplet detachment on a convex surface. Stages I-II schematically illustrate a droplet spreading on a curved surface due to the EW effect. Recoiling and detachment of a droplet after the switch-off of the applied voltage on a curved surface are illustrated as stages III-V.\label{schematic}}
\end{figure}

We begin our investigation by considering a water droplet placed on a curved solid substrate (curvature $1/R$) consisting of an insulating surface layer (thickness $d$) on top (purple) and an electrode underneath, as shown in Fig.~\ref{schematic}(I). The apparent contact angle $\theta_{\rm Y}$ at the equilibrium state satisfies Young's equation in the absence of voltage, $\gamma_{\rm sm}-\gamma_{\rm ls}=\gamma_{\rm lm}{\rm cos}\theta_{\rm Y}$, where $\gamma_{\rm sm}$ and $\gamma_{\rm ls}$ are interfacial tensions of the solid-medium and liquid-solid interfaces, respectively.
When an external voltage $U$ is applied between a droplet and a flat substrate, the accumulation of free charges near the electrode causes a reduction of the local liquid-solid surface tension $\gamma_{\rm ls}$ and subsequently induces the spreading of the droplet, yielding another equilibrium state, with its new equilibrium contact angle $\theta_{\rm e}$ described by the well-known Young-Lippmann equation~\cite{Lippmann1875,F.Mugele2005} ${\rm cos}~ \theta_{\rm e} = {\rm cos}~ \theta_{\rm Y} + \varepsilon_0 \varepsilon U^2/2d\gamma_{\rm lm}$. For a spherical curved surface, as shown in Fig.~\ref{schematic}(II), this equation has to be replaced by~\cite{Y.Wang2012}
\begin{align}
{\rm cos}~ \theta_{\rm e} = {\rm cos}~ \theta_{\rm Y} + \frac{\varepsilon_0 \varepsilon U^2}{2d\gamma_{\rm lm}}\cdot \frac{1}{1\pm\kappa}={\rm cos}~ \theta_{\rm Y}+\eta,\label{eq:Y-L}
\end{align}
where $\kappa=d/R$ represents the effect of surface curvature, and $\varepsilon_{0}$, $\varepsilon$, $\gamma_{\rm lm}$ and $\eta$ are the dielectric permittivity in vacuum, the relative dielectric constant, the interfacial tension between the droplet and its surrounding medium above the substrate, and the dimensionless EW number, respectively. We use the sign $\pm$ to distinguish the effect of convex and concave surfaces here and throughout the rest of this paper, with $+$ standing for convex surface and $-$ for concave one.
Once the applied electric voltage is switched off [Fig.~\ref{schematic}(III)], the droplet will keep its current shape as that of stage II as its transient state, but with a sudden increase of the local liquid-solid surface tension owing to the short time scale of discharge process of the droplet-electrode capacitor, which is much faster than the relaxation time of the droplet~\cite{A.Cavalli2016,K.X.Zhang2019}. As a consequence, the restoration of liquid-solid interfacial tension after the switch-off of external voltage increases the surface energy of the droplet-substrate system instantly, leading to a total free interfacial energy expressed as
\begin{align}
E_{\rm s}^{\rm III}= \gamma_{\rm lm} A_{\rm c} \biggl( \frac{1+{\cos}\Delta\theta}{1+{\cos}\theta}-{\cos}\theta_{\rm Y} \biggr).\label{eq2:ESIII.2}
\end{align}
where $\Delta\theta=\mid\theta-\theta_{\rm e}\mid$, with $\theta$ the apparent contact angle, and $A_{\rm c}=2 \pi R^2(1-{\rm cos}\Delta\theta)$ is the contact area under the spherical droplet cap. Here it is necessary to note that Young's equation has been used in deriving the above equation.

Subsequently, as a result of recovery effect, the droplet undergoes a recoiling stage [Fig.~\ref{schematic}(IV)], during which the flow of droplet will induce a viscous dissipation typically comprising of contributions in the bulk droplet, near the substrate, and at the vicinity of the contact line~\cite{Q.Vo2018SP,Q.Vo2018PRE,Q.Vo2019}. As a result, the viscous dissipation for the bulk flow and that near the substrate can be estimated via $E_{\rm vis}^{\rm b,s}=\int_{0}^{\tau}\int_{\rm V}\Phi dVdt$~\cite{S.Chandra1991}. Meanwhile, the energy of dissipation contributed by the contact line can be approximately calculated as
\begin{align}
E_{\rm vis}^{\rm c} = \gamma_{\rm lm} A_{\rm c} \frac{\lambda Oh R\Delta\theta\eta^{1/2}}{\pi \mu R_{0}},\label{eq:Evisc}
\end{align}
where $\lambda$, $\mu$, $R_0$, and $Oh=\mu/\sqrt{\rho\gamma_{\rm lm}R_0}$ are, respectively, the friction coefficient at contact line, the viscosity of the droplet, the initial radius of the droplet, and the Ohnesorge number, which describes the relative importance of viscosity and surface tension. Here parameter $\rho$ in $Oh$ denotes the density of the droplet. Strictly speaking, the total viscous dissipation is the sum of $E_{\rm vis}^{\rm b,s}$ and $E_{\rm vis}^{\rm c}$. However, as the contribution at the vicinity of the contact line dominates the whole dissipation process in the system~\cite{Q.Vo2019}, we therefore merely consider the contribution made by the contact line.
In addition, apart from the viscous dissipation, the pinning and subsequently the stretching of the liquid interface at the contact line vicinity during the recoiling process also lead to an accumulation of energy~\cite{Q.Vo2019} estimated as
\begin{align}
E_{\rm stored} = \gamma_{\rm lm} A_{\rm c} \frac{\pi{\rm sin}^2\theta_{\rm r}}{\ln(L\sigma^{-1})},\label{eq:Estored}
\end{align}
where $\theta_{\rm r}$, $L$, and $\sigma$ are the receding contact angle, the macroscopic cutoff length, and the topological defect size of the substrate, respectively. Here the expression on a spherical surface $\theta_{\rm r}=\theta_{\rm r,0}\pm\Delta\theta$, instead, is used for the receding contact angle~\cite{C.W.Extrand2008}, with $\theta_{\rm r,0}$ the droplet's intrinsic receding contact angle on curved surface.
Finally, at the end of the retracting stage, if the total free interfacial energy of stage III is sufficient to overcome the energy barrier hindering droplet jumping, the droplet will detach from the substrate, as illustrated in Fig.~\ref{schematic}(V), corresponding to a total energy written as
\begin{align}
E_{\rm s}^{\rm V} =4\pi R_0^2\gamma_{\rm lm}+E_{\rm k},\label{eq:EV}
\end{align}
where the first term $E_{\rm s}^{\rm V}=4\pi R_0^2\gamma_{\rm lm}$ and the second one are the surface energy and the kinetic energy of the droplet, respectively. $E_{\rm k}>0$ means detachable and $E_{\rm k}<0$ implies nondetachable, and the critical condition $E_{\rm k}=0$ for the detachment corresponds to
\begin{align}
\Delta E_{\rm s} =E_{\rm s}^{\rm III}-E_{\rm s}^{\rm V}= E_{\rm vis}^{\rm c}+E_{\rm stored},\label{eq:criticalsondition}
\end{align}
where $\Delta E_{\rm s}$ denotes the surface energy difference between stages III and V.
Here we assume that the droplet volume is small, corresponding to an initial radius smaller than the capillary length $l_{\rm c}=\sqrt{\gamma_{\rm lm}/\rho g}\sim1.9~{\rm mm}$. Thus the gravitational effect can be neglected, and the droplet at stages III and V can be treated as a spherical one in this paper. In addition, the Marangoni and the evaporation effects of the aqueous droplet are excluded as well, thereby a reasonable assumption is made that the volume of the droplet $V_0$ is conserved during the whole process, corresponding to a constraint
\begin{align}
V_0=\frac{\pi}{3}R^3\frac{{\sin}^3\Delta\theta}{{\sin}^3\theta}f(\theta)-\biggl[ \pm \frac{\pi}{3}R^3f(\Delta\theta)\biggr]=\frac{4\pi}{3}R_0^3,\label{eq:volumeconservation}
\end{align}
where $f(\theta)=2-3{\cos}\theta +{\cos}^3\theta$ is a dimensionless function.
Here it is reasonable to assume that $\Delta\theta$ is a small quantity when $R\gg R_0$, hence we can do a Taylor expansion with respect to $\Delta\theta$ for Eq.~(\ref{eq:volumeconservation}), leading to
\begin{align}
\Delta\theta=\biggl[\frac{4}{f(\theta_{\rm e})}\biggr]^{1/3}\frac{R_0}{R}{\sin}\theta_{\rm e}.\label{eq:deltatheta}
\end{align}
In order to investigate the influence of surface curvature on the critical detachment condition of the droplet, we substitute Eqs.~(\ref{eq2:ESIII.2}),~(\ref{eq:Evisc}),~(\ref{eq:Estored}), and~(\ref{eq:EV}) into Eq.~(\ref{eq:criticalsondition}) and do the same Taylor expansion for both sides. By using Eq.~(\ref{eq:deltatheta}), the critical condition for the detachment of droplet on a curved surface can be derived as
\begin{align}
\Gamma-\Omega-\frac{\pi{\sin}^2\theta_{\rm r,0}}{\ln(L\sigma^{-1})}=\pm\biggl[\frac{\pi{\sin}(2\theta_{\rm r,0})}{\ln(L\sigma^{-1})}-\frac{2{\sin}\theta_{\rm e}}{(1+{\cos}\theta_{\rm e})^2}\biggr]\Delta\theta,\label{eq:criticaldetachmentcondition}
\end{align}
where $\Gamma=2/(1+{\cos}\theta_{\rm e})-4\bigl[4/f(\theta_{\rm e})\bigr]^{-2/3}{\sin}^{-2}\theta_{\rm e}-{\cos}\theta_{\rm Y}$ represents contribution of excessive surface energy, and $\Omega=(\lambda Oh/\pi\mu)\eta^{1/2}\bigl[4/f(\theta_{\rm e})\bigr]^{1/3}{\sin}\theta_{\rm e}$ denotes the energy of viscous dissipation, both depending on surface curvature due to Eq.~(\ref{eq:Y-L}).
Intriguingly, the critical detachment condition of droplet on a curved surface Eq.~(\ref{eq:criticaldetachmentcondition}) returns exactly to the one on a flat surface~\cite{Q.Vo2019} if the surface curvature $1/R$ approaches zero.

\begin{figure}[htp]
%\centerline{\includegraphics[width=1.0\textwidth,keepaspectratio]{figure2}}
  \includegraphics[width=\linewidth,keepaspectratio]{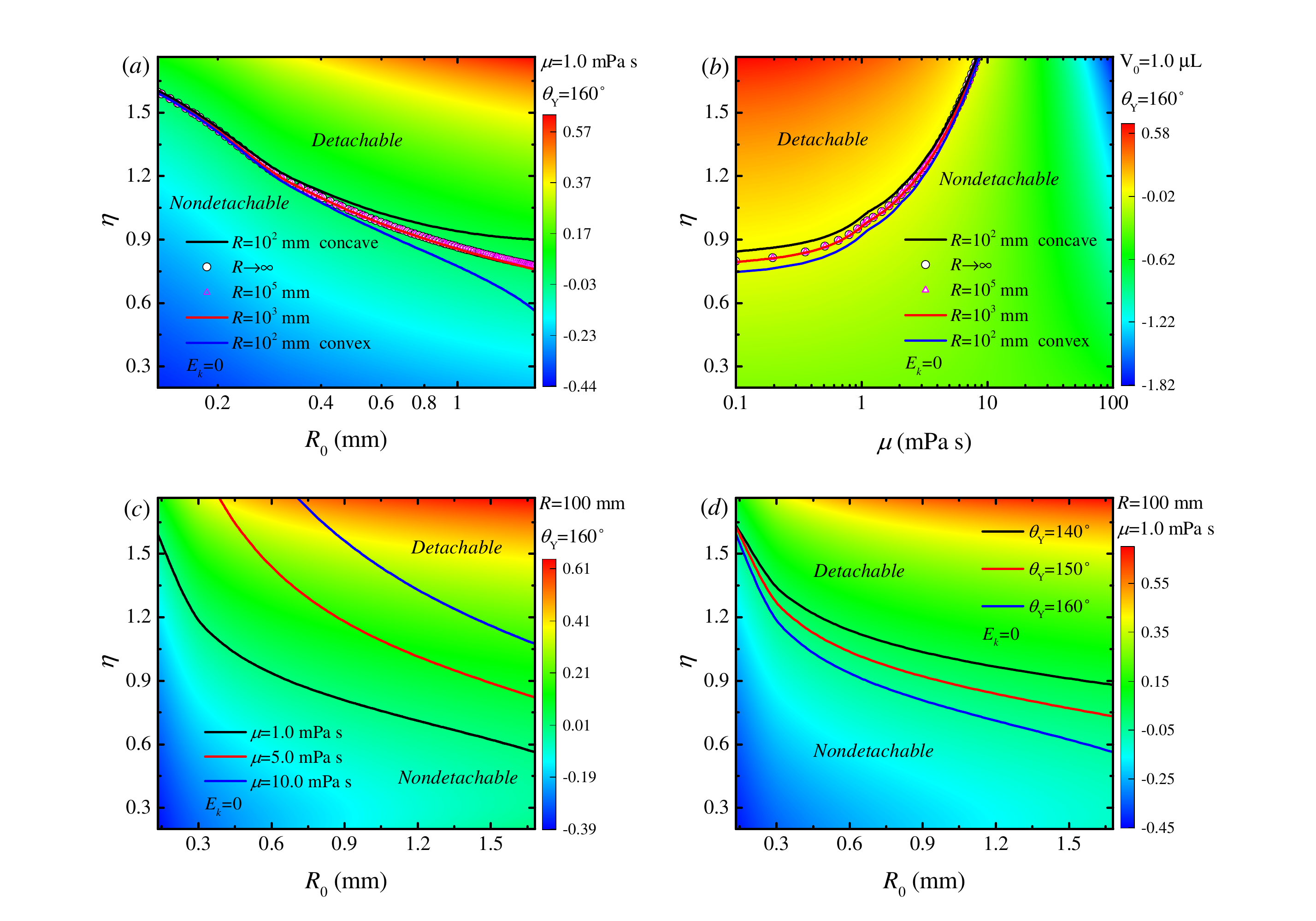}
  \caption{(Color online) Phase diagrams for the detachment of droplet in the projection planes of (a) electrowetting number $\eta$ and droplet radius $R_0$ with droplet viscosity $\mu$=1.0 mPa~s; (b) electrowetting number $\eta$ and droplet viscosity $\mu$ with droplet volume $V_0=1.0~{\rm \mu L}$, where the Young's contact angle is set as $\theta_{\rm Y}=160^{\circ}$, and the curve lines denote the critical detachment conditions for different surface radii ($\infty$, $10^5$, $10^3$, and $10^2~{\rm mm}$). Phase diagrams in ($\eta$, $R_0$) plane for surface radius $R=10^2~{\rm mm}$, where the curve lines denote the critical droplet detachment conditions for (c) different droplet viscosities (1.0, 5.0, and 10.0 mPa s) with $\theta_{\rm Y}=160^{\circ}$, and (d) different surface wettabilities ($\theta_{\rm Y}=140^{\circ}, 150^{\circ}$, and $160^{\circ}$) with droplet viscosity $\mu$=1.0 mPa s. \label{RoandMuVsEta}}
\end{figure}
Our calculation is carried out by using the same values of parameters as those in Ref.~\cite{Q.Vo2019}, i.e., $\gamma_{\rm lm}=37.2~{\rm mN \cdot m^{-1}}$, $\rho=1.003~{\rm g\cdot cm^{-3}}$, $\varepsilon=1.93$, $d=2.2~{\rm\mu m}$, $\theta_{\rm r,0}=121.7^{\circ}$, $\sigma=464~{\rm nm}$ and $\lambda=C(\mu\mu_0)^{1/2}$, where $C$ and $\mu_0$ (viscosity of the surrounding medium of droplet) are fixed as 34 and 1.8 mPa s.
In order to reveal the effect of droplet volume (which has been converted to the radius of droplet $R_0$), liquid viscosity, and EW number $\eta$ on droplet detachment, we constructed phase diagrams in $\eta-R_0$ (EW number versus radius) and $\eta-\mu$ (EW number versus viscosity) space, as demonstrated in Fig.~\ref{RoandMuVsEta}.
It is found that all phase diagrams are divided into two regimes, namely nondetachable phase regime and detachable phase regime separated by a coexisting line representing the critical condition for a droplet to detach from a curved surface, above which the excess surface energy will be converted into the kinetic energy of the detached droplet. A specific case of our model corresponding to a droplet on a flat surface($R\rightarrow \infty$) is also shown (white circle) in Figs.~\ref{RoandMuVsEta}(a) and~\ref{RoandMuVsEta}(b). On the one hand, for a fixed surface radius, i.e., $R=10^3 ~{\rm mm}$ (red critical line in Figs.~\ref{RoandMuVsEta}(a) and~\ref{RoandMuVsEta}(b)), increasing droplet radius $R_0$ or decreasing droplet viscosity $\mu$ clearly reduces the detachment threshold value of voltage applied, indicating that, for a small droplet with high viscosity, a large excessive surface energy is needed to overcome the viscous dissipation before detachment, which correspondingly requires a higher electric voltage applied.
On the other hand, Figs.~\ref{RoandMuVsEta}(a) and~\ref{RoandMuVsEta}(b) also show that surface curvature plays a significant role in deciding the threshold value of the electric voltage applied for the detachment to occur. If we increase the surface curvature to $10^{-2}$ mm$^{-1}$, but still small compared to that of the droplet $1/R_0$, the threshold value of the applied electric voltage for detachment significantly decreases for convex surface as compared with that of flat surface, as shown by the blue curve line in Figs.~\ref{RoandMuVsEta}(a) and~\ref{RoandMuVsEta}(b). However, a higher threshold value of the applied electric voltage is required for the occurrence of droplet detachment on concave surfaces, as depicted by the black line in Figs.~\ref{RoandMuVsEta}(a) and~\ref{RoandMuVsEta}(b).
\begin{figure}[htp]
  \includegraphics[width=\linewidth,keepaspectratio]{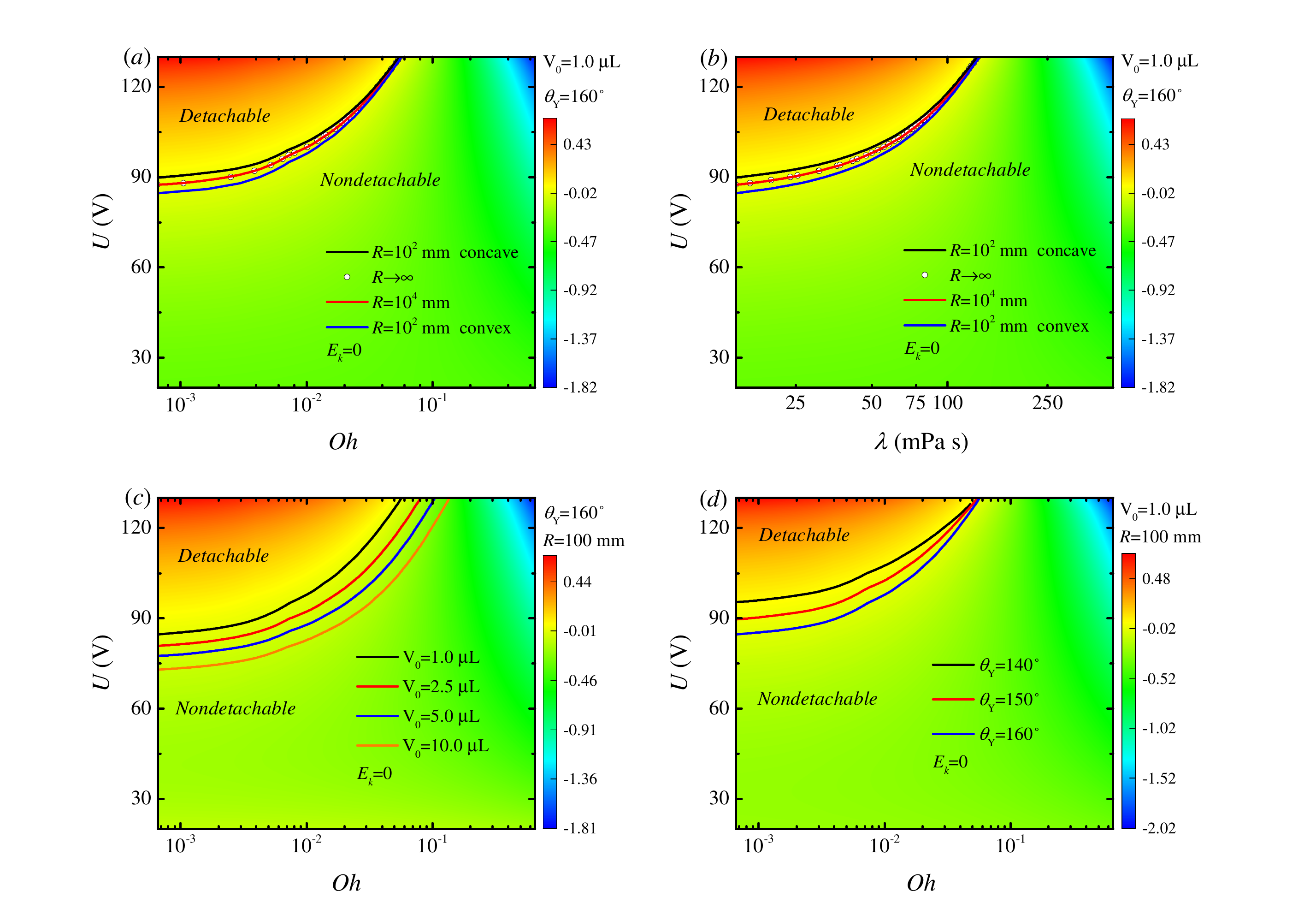}
  \caption{(Color online)
  Phase diagrams of the applied electric voltage $U$ vs (a) Ohnesorge number $Oh$ and vs (b) friction coefficient $\lambda$, with droplet volume $V_0$=1.0~$\mu$L and the Young's contact angle $\theta_{\rm Y}=160^{\circ}$, where the curve lines denote the critical droplet detachment conditions for different surface radii ($\infty$, $10^4$, and $10^2~{\rm mm}$). Phase diagrams $U$ vs $Oh$ for surface radius $R=10^2~{\rm mm}$, where the curve lines denote the critical droplet detachment conditions for (c) different droplet volumes (1.0, 2.5, 5.0, and 10.0~$\mu {\rm L}$) with $\theta_{\rm Y}=160^{\circ}$, and (d) different surface wettabilities ($\theta_{\rm Y}=140^{\circ}, 150^{\circ}$, and $160^{\circ}$) with droplet volume $V_0$=1.0~$\mu$L. \label{OhandLambdaVsU}}
\end{figure}
Meanwhile, the critical curve lines in $\eta-R_0$ space for different values of droplet viscosity and Young's contact angle are presented in Figs.~\ref{RoandMuVsEta}(c) and~\ref{RoandMuVsEta}(d), where it is found that detachable regime is remarkably shrunk as we increase the droplet viscosity and reduce the surface wettability, a result in agreement with the conclusion drawn from Fig.~\ref{RoandMuVsEta}(b).

As it is clearly seen that, according to Eq.~\ref{eq:Evisc}, the energy of viscous dissipation is directly related to Ohnesorge number and the friction coefficient at contact line during the recoiling process, it is meaningful to probe the influence of Ohnesorge number $Oh$ and friction coefficient $\lambda$ on the critical voltage value. Here accordingly we plot the phase diagram of detachment projected in $U-Oh$ (applied electric voltage versus Ohnesorge number) plane and $U-\lambda$ (applied electric voltage versus friction coefficient) plane, as shown in Fig.~\ref{OhandLambdaVsU}. For a fixed set of the parameters (i.e., surface radius $R=\infty$, droplet volume $V_0=1.0~\mu{\rm L}$, and Young's contact angle $\theta_{\rm Y}=160^{\circ}$), Figs.~\ref{OhandLambdaVsU}(a) and~\ref{OhandLambdaVsU}(b) demonstrate that a higher electric voltage is required to detach the droplet from the substrate with the increase of Ohnesorge number and the friction coefficient.
Here it is worthwhile to note that within some fixed parameters range, droplet detachment is unable to occur at large values of $Oh$ and $\lambda$~\cite{Q.Vo2019} because the excessive surface energy in this case is largely dissipated, except for a small detachable regime in parameter space restricted at the left top corner of the figure where $Oh$, and $\lambda$ are small.
Such a feature does not depend on the curvature of the substrate when the surface curvature is not small enough, i.e. the red curve with $R=10^4 ~{\rm mm}$ overlaps that for flat surface (white circle with $R\rightarrow\infty$), as shown in Figs.~\ref{OhandLambdaVsU}(a) and~\ref{OhandLambdaVsU}(b). However, for the black (concave) and blue (convex) curves with $R=10^2 ~{\rm mm}$, we still see deviations to different directions from that of Figs.~\ref{RoandMuVsEta}(a) and~\ref{RoandMuVsEta}(b), indicating that it is more difficult to detach a droplet from a concave substrate in comparison with a convex one.
Furthermore, the critical curve line in $U-Oh$ space under different droplet volumes (1.0, 2.5, 5.0, and 10.0~$\mu$L) and Young's contact angles ($140^{\circ}$, $150^{\circ}$, and $160^{\circ}$) are plotted in Figs.~\ref{OhandLambdaVsU}(c) and~\ref{OhandLambdaVsU}(d), where the variation of the detachable regime once
\begin{figure}
  \includegraphics[width=\linewidth]{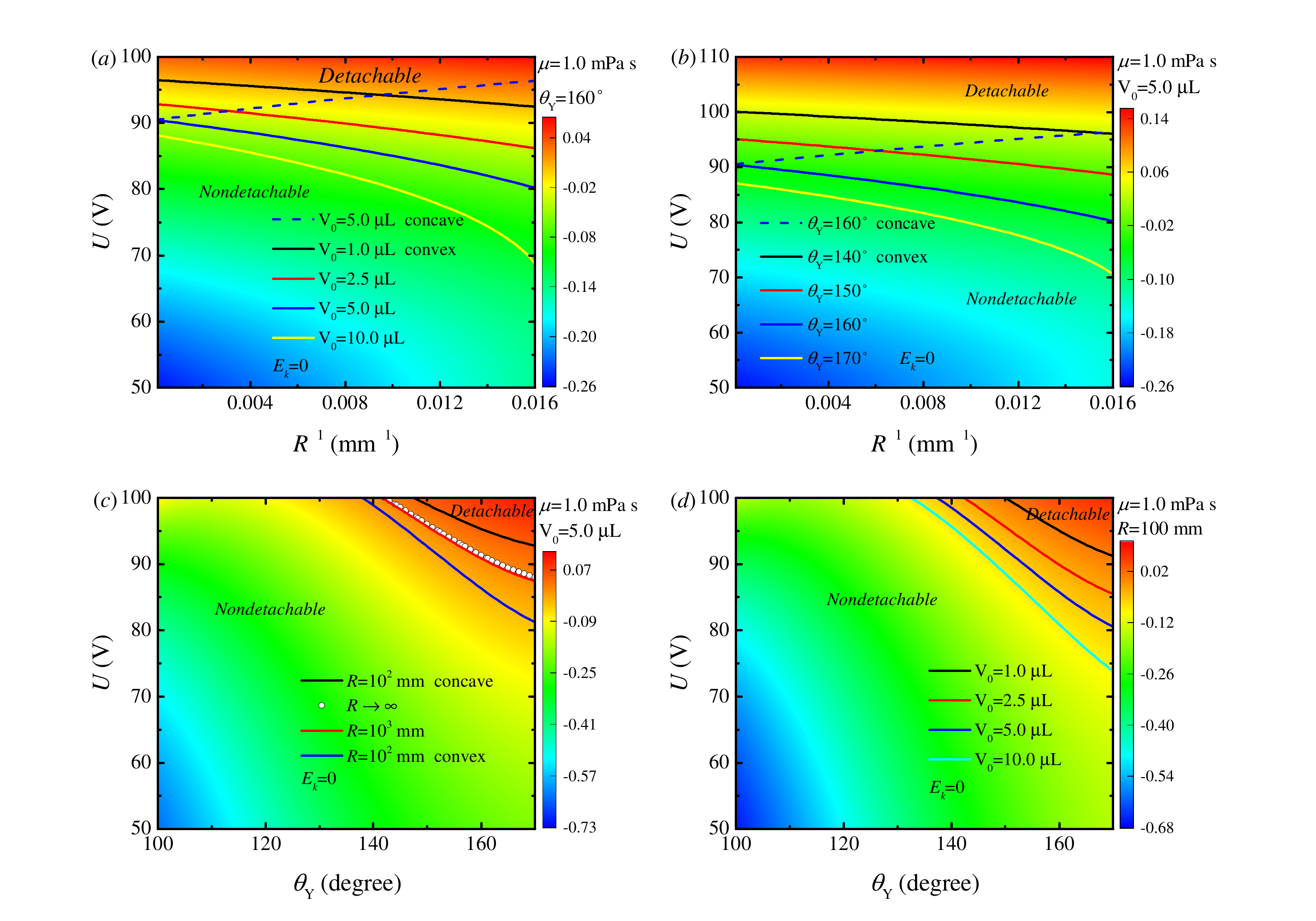}
  \caption{(Color online) Phase diagrams show detachable and nondetachable regimes in the parameter space spanned by the applied electric voltage $U$ and the surface curvature $R^{-1}$ with droplet viscosity $\mu$=1.0 mPa s, where the curve lines denote the critical droplet detachment conditions for (a) different droplet volumes (1.0, 2.5, 5.0, and 10.0~$\mu {\rm L}$) with $\theta_{\rm Y}=160^{\circ}$, and (b) different surface wettabilities ($\theta_{\rm Y}=140^{\circ}, 150^{\circ}$, and $160^{\circ}$) with droplet volume $V_0=5.0~\mu {\rm L}$. Similar phase diagrams spanned by the applied electric voltage $U$ and the Young's contact angle $\theta_{\rm Y}$ with droplet viscosity $\mu$=1.0 mPa s, where the curve lines denote the critical droplet detachment conditions for (c) different surface radii ($\infty$, $10^3$, and $10^2~{\rm mm}$) with droplet volume $V_0=5.0~\mu {\rm L}$, and (d) different droplet volumes (1.0, 2.5, 5.0, and 10.0~$\mu {\rm L}$) with surface radius $R=10^2~{\rm mm}$. \label{CurvatureandThetaYVsU}}
\end{figure}
again verifies that smaller droplet and less hydrophobic surface have a negative effect on droplet detachment, and thus require a higher critical electric voltage to induce detachment.

Finally, in order to understand how surface curvature and surface wettability affect the critical value of applied electric voltage, phase diagrams of detachment in $U-R^{-1}$ (applied voltage versus surface curvature) space and $U-\theta_{\rm Y}$ (applied voltage versus Young's contact angle) space are constructed, as shown in Fig.~\ref{CurvatureandThetaYVsU}.
Interestingly, the critical curve lines for concave surface (blue dashed line) and convex surface (blue solid line) in Figs.~\ref{CurvatureandThetaYVsU}(a) and~\ref{CurvatureandThetaYVsU}(b) show almost opposite behaviors in response to the variation of surface curvature, that is, the critical electric voltage required for detachment decreases with the increase of $R^{-1}$ for convex surface, while higher critical electric voltage for detachment is needed as we increase $R^{-1}$ for concave surface. Meanwhile,the detachable regimes in parameter space are also apparently broadened with the increase of droplet volume and Young's contact angle [see Figs.~\ref{CurvatureandThetaYVsU}(a) and~\ref{CurvatureandThetaYVsU}(b)].
In addition, the critical curve in $U-\theta_{\rm Y}$ space for different surface radii ($\infty, 10^3$, and $10^2$~mm) and droplet volumes (1.0, 2.5, 5.0, and 10.0~$\mu$L) are presented in Figs.~\ref{CurvatureandThetaYVsU}(c) and~\ref{CurvatureandThetaYVsU}(d), where we can deduce that the decrease of wettability, namely increasing Young's contact angle (hydrophobicity), will lead to a decrease of the threshold value of the electric voltage applied to the droplet for detachment.
It is also found from the figure that the red curve for curved surface ($R=10^3$~mm) overlaps the circle curve for flat surface ($R\rightarrow\infty$), indicating that the surface curvature is not large enough to bring about a change in critical electric voltage for the detachment to occur. However, the detachable regime is dramatically widened (narrowed down) for convex (concave) surface at $R=10^2$~mm (still large enough as compared with $R_0$) if compared with that for flat surface, as shown by the blue line and black line in Fig.~\ref{CurvatureandThetaYVsU}(c).
Furthermore, the variation tendency of the critical curves in $U-\theta_{\rm Y}$ space as shown in Fig.~\ref{CurvatureandThetaYVsU}(d) denotes that the detachable regime is also affected by the size of droplet, or more specifically, the detachable regime is noticeably broadened with the increase of droplet volume. As a result, Fig.~\ref{CurvatureandThetaYVsU} confirms that large droplet and superhydrophobicity of surface both play a positive role in inducing a droplet detachment.

Consequently, we argue that larger droplet volume, lower droplet viscosity, Ohnesorge number, the friction coefficient, large (convex surface) or small (concave surface) surface curvature and high hydrophobicity favors droplet detachment.
Since whether the detachment occurs or not largely relies on the competition among the three types of energy, namely the excessive surface energy, the energy of viscous dissipation, and the accumulated energy at the vicinity of the contact line.
The droplet detachment only occurs under the condition that the excessive surface energy is sufficient to overcome the energy barrier, namely the sum of viscous dissipation and the accumulated energy around the contact line. In order to examine the validity of our theoretical model, it is necessary to compare our theoretical predictions with experimental results and other theoretical analyses.
For example, it has been reported both experimentally~\cite{A.Cavalli2016} and theoretically~\cite{K.X.Zhang2019} that increasing Young's contact angle gives rise to an increase in EW-induced velocity of the detached droplet, a result in good agreement with our present arguments.
By employing high density ratio-based lattice Boltzmann method, Raman et al.~\cite{K.A.Raman2016} have examined the effect of Ohnesorge number on the jumping velocity of the detached droplet and demonstrated that increasing Ohnesorge number leads to a decrease of droplet jumping velocity, which is also in accordance with our conclusions. In addition, it has been shown, by providing an analytical method to study EW-induced jumping of droplet on flat hydrophobic substrates, that jumping motion of droplet can be enhanced via increasing Young's contact angle or decreasing Ohnesorge number~\cite{K.X.Zhang2019}, a conclusion which can be made and has been included by our theoretical model. What's more, Wang et al.~\cite{Q.G.Wang2020} showed via experiments that higher threshold voltage is needed to induce detachment of droplet with larger friction coefficient and smaller volume. These results all support the theoretical model we present in this paper.

In summary, we investigate the detachment of an aqueous droplet induced by EW effect on a curved surface, and derive a general criterion equation for the detachment to occur. According to the obtained phase diagrams consisting of detachable and nondetachable regimes, we show that the direct droplet detachment can be triggered by applying an electric voltage (corresponding to EW number) beyond a critical value and then switching it off. It is found that, by judging the critical curve of voltage applied, it is easier (more difficult) for a droplet on a convex (concave) surface to detach if compared with one on a flat surface. The dependence of the threshold value of the applied voltage on droplet volume, liquid properties, surface curvature, and Young's contact angle is discussed. In particular, the results here suggest that the threshold value of the applied electric voltage decrease with the increase of droplet volume, surface curvature (convex surface) and Young's contact angle, or the decrease of droplet viscosity, Ohnesorge number, friction coefficient and surface curvature (concave surface). Therefore, it is possible to harness the detachment of a droplet triggered by switching off the electric voltage applied, by tuning the volume and the viscosity of the droplet, and the wettability and the curvature of the substrate. Here it should be noted that the gravitational effect and the oscillation of the droplet are both ignored in our study.

\begin{acknowledgements}
This work was funded by the National Science Foundation of China under Grant No. 11974292 and No. 11947401.
\end{acknowledgements}

%\bibliography{swimmer} %You need to replace "rsc" on this line with the name of your .bib file

\end{document}